\documentclass[11pt,twoside]{article}
\usepackage{asp2010}

\resetcounters

\begin{document}

\title{Magnetic fields around (post-)AGB stars and (Pre-)Planetary Nebulae}
\author{W.~H.~T.~Vlemmings
\affil{Argelander-Institut f{\"u}r Astronomie, University of Bonn, Auf dem H{\"u}gel 71, D-53121 Bonn, Germany}}

\begin{abstract}
  Observational evidence for strong magnetic fields throughout the
  envelopes of evolved stars is increasing. Many of the
  instruments coming on line in the near-future will be able to make
  further contributions to this field. Specifically, maser
  polarization observations and dust/line polarization in the sub-mm
  regime has the potential to finally provide a definite picture of
  the magnetic field strength and configuration from the Asymptotic
  Giant Branch (AGB) all the way to the Planetary Nebula
  phase. While current observations are limited in sample size, strong
  magnetic fields appear ubiquitous at all stages of (post-)AGB
  evolution. Recent observations also strongly support a field
  structure that is maintained from close to the star to several
  thousands of AU distance. While its origin is still unclear, the
  magnetic field is thus a strong candidate for shaping the stellar
  outflows on the path to the planetary nebula phase and might even
  play a role in determining the stellar mass-loss.\\
  \noindent{\bf Keywords.}\hspace{10pt}Planetary Nebulae -- Stars: AGB and post-AGB -- Magnetic fields
\end{abstract}

\section{Introduction}

Strongly asymmetric planetary nebulae (PNe) have been shown to be
common. The research into their shaping processes has become a fundamental part of our
attempts to further the understanding of the return of processed
material into the ISM by low- and intermediate-mass stars at the end
of their evolution. Whereas the standard interacting winds scenario
\citep{Kwok78} can explain a number of the PN properties, an important
discovery has been that the collimated outflows of the pre-PNe
(P-PNe), where such outflows are common, have a momentum that exceeds
that which can be supplied by radiation pressure alone
\citep{Bujarrabal01}. The source of this momentum excess has been
heavily debated during the past several years, with the most commonly
invoked cause being magnetic fields, binary or disk interaction or a
combination of these \citep[e.g.][]{Balick02}. Due to a number of
similarities with the jets and outflows produces by young stellar
objects, the study of P-PNe outflows provides further research
opportunities into a potentially universal mechanism of jet launching.

Here I will review the observational evidence for strong magnetic
fields in PNe as well as around their AGB and post-AGB progenitors.  I
will give an overview of the methods that can be used to study
magnetic fields, especially in light of the plethora of new
instruments that will be available shortly. Finally, I will discuss a
number of questions related this topic that we can expect to be
answered with the new instruments in the next few years.

\section{Observational Techniques - Polarization}

With the exception of observations where the magnetic field strength
is estimated assuming forms of energy equilibrium, such as synchrotron
observations, the magnetic field strength and structure is typically
determined from polarization observations. 

\subsection{Circular Polarization}

Circular polarization, generated through Zeeman splitting, can be used
to measure the magnetic field strength. It measures the total field
strength when the splitting is large and the line-of-sight component
of the field when the splitting is small. The predominant source of
magnetic field strength information during the late stages of stellar
evolution comes from maser circular polarization observations, and
particularly the common SiO, H$_2$O and OH masers. These can show
circular polarization fractions ranging from $\sim0.1$\% (H$_2$O) up
to $\sim100$\% (OH) and are, because of their compactness and
strength, excellent sources to be observed with high angular
resolution. Unfortunately, the analysis of maser polarization is not
straightforward \citep[For a review, see][]{Vlemmings07}, and it has
taken a long time before maser observations were acknowledged to
provide accurate magnetic field measurements. More recently, the first
attempts have been made to detect the Zeeman splitting of non-maser
molecular lines in circumstellar envelopes, such as CN
\citep{Herpin09}. As many of these occur at shorter wavelength in the
(sub-)mm regime, the advent of the Atacama Large (sub-)Millimeter
Array will further enhance these types of studies.

\subsection{Linear Polarization}

Linear polarization, probing the structure of the plane-of-the-sky
component of the magnetic field, can be observed both in the dust
(through aligned grains) and molecular lines (through radiation
anisotropy - the Goldreich-Kylafis effect). Typical percentages of
linear polarization range from up to a few percent (e.g. dust, CO,
H$_2$O masers) to several tens of percent (OH and SiO masers). Again
the interpretation of maser polarization depends on a number of
intrinsic maser properties, but in specific instances maser linear
polarization can even be used to determine the full 3-dimensional
field morphology. In addition to the geometry, the linear polarization
of most notably dust, can also be used to obtain a value for the
strength of the plane-of-the-sky component of the magnetic
field. This is done using the Chandrasekhar-Fermi method, which refers to the
relation between the turbulence induced scatter of polarization
vectors and the magnetic field strength.

\section{Current Status - Evolved Star Magnetic Fields}

\subsection{AGB Stars}

\articlefigure[width=10cm]{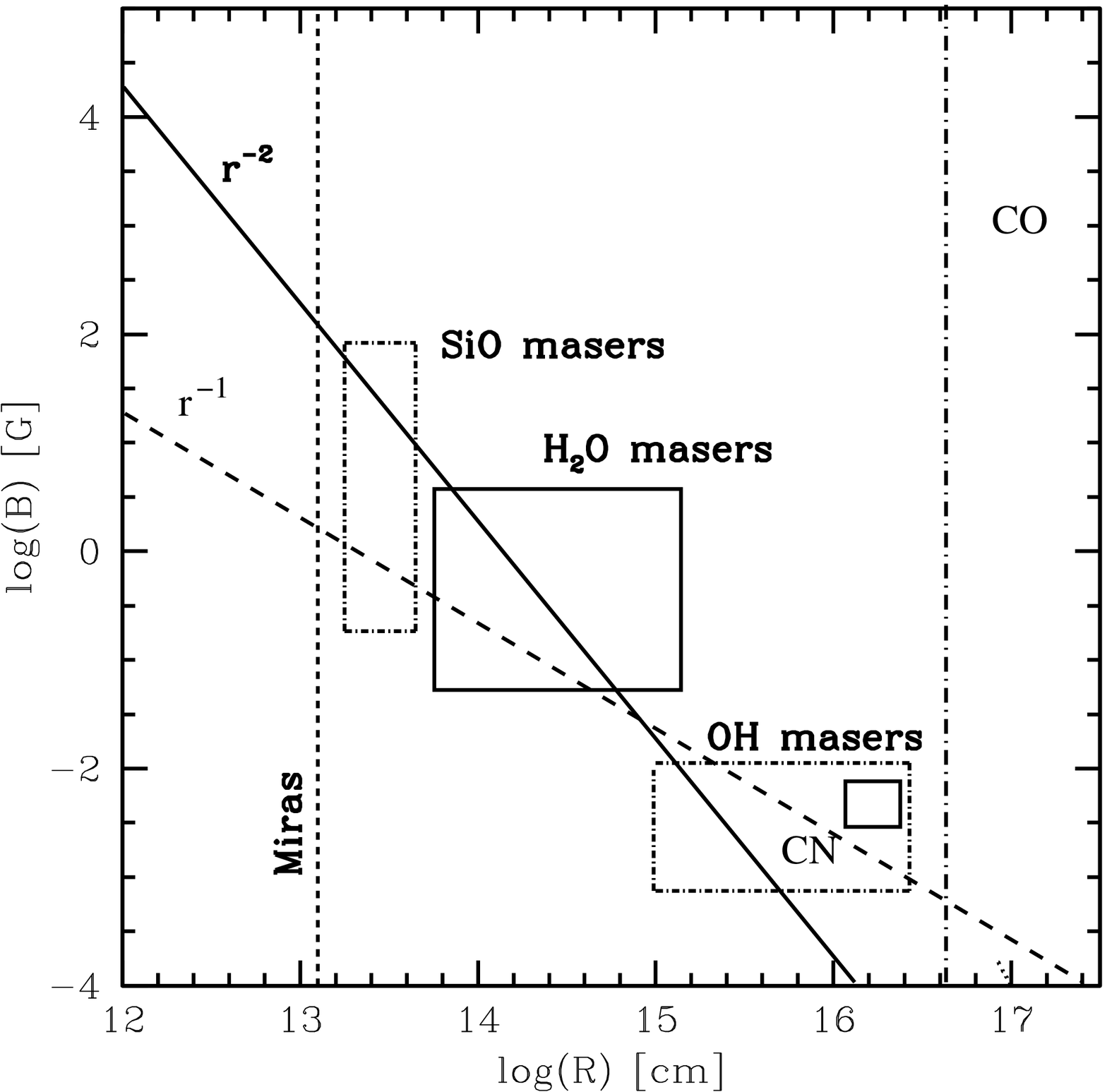}{bvsr}{Magnetic field
  strength vs. radius relation as indicated by current maser
  polarization observation of a number of Mira stars. The boxes show
  the range of observed magnetic field strengths derived from the
  observations of SiO masers \citep{Kemball09, Herpin06}, H$_2$O masers
  \citep{Vlemmings02, Vlemmings05}, OH masers
  \citep[e.g.][]{Rudnitski10} and CN \citep{Herpin09}. The thick solid
  and dashed lines indicate an $r^{-2}$ solar-type and $r^{-1}$
  toroidal magnetic field configuration. The vertical dashed line
  indicates the stellar surface. CO polarization observations will
  uniquely probe the outer edge of the envelope (vertical dashed dotted
  line).}

\articlefigure[width=14cm]{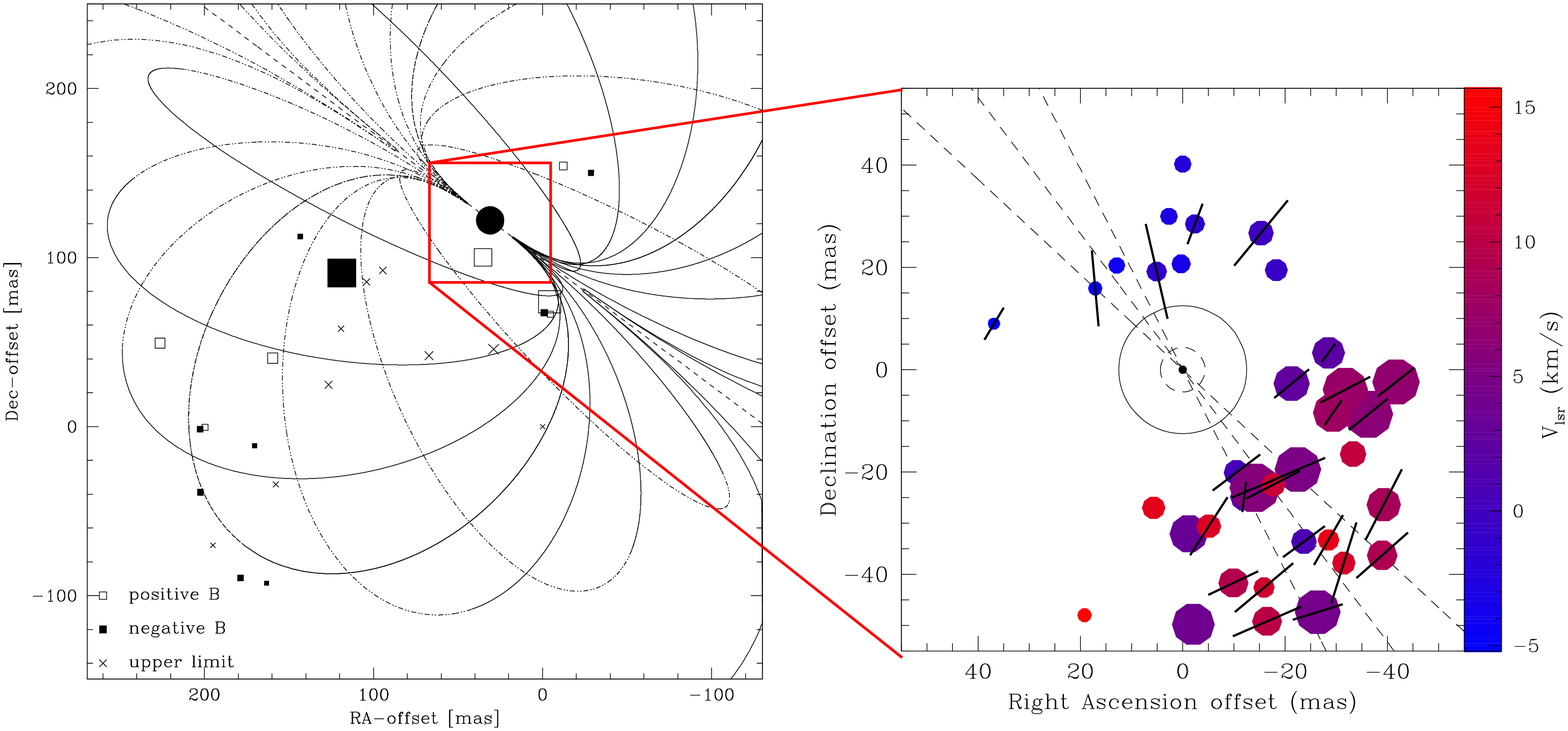}{vxsgr}{ (left) The dipole
  magnetic field of the supergiant VX~Sgr as determined from a fit to
  the H$_2$O maser magnetic field observations
  \citep{Vlemmings05}. (right) Positions and polarization of the
  VX~Sgr $v=0, J=5-4$ $^{29}$SiO masers observed with the SMA
  \citep{Vlemmings10}. The masers spots are plotted with respect to
  the peak of the continuum emission. The black vectors are the
  observed polarization vectors scaled linearly according to
  polarization fraction. The long dashed inner circle indicates the
  star and the solid circle indicates the location of the 43~GHz SiO
  masers. The short dashed circle indicates the minimum radius of the
  $^{28}$SiO masers. The dashed lines indicate the position angle and
  its uncertainty of the inferred orientation of the dipole magnetic
  field of VX~Sgr observed using H$_2$O and OH
  masers\citep{Vlemmings05, Szymczak01}.}

Most AGB magnetic field measurements come from maser polarization
observations (SiO, H$_2$O and OH). These have revealed a strong
magnetic field throughout the circumstellar envelope. In
Figure~\ref{bvsr}, I have indicated the magnetic field strength in the
regions of the envelope traced by the maser measurements throughout
AGB envelopes. While a clear trend with increasing distance from the
star is seen, the lack of accurate information on the location of the
maser with respect to the central stars makes it difficult to
constrain this relation beyond stating that it seems to vary between
$B\propto R^{-2}$ (solar-type) and $B\propto R^{-1}$
(toroidal). Future observations of CO polarization might be able to
provide further constraints.

As the masers used for these studies are mostly found in oxygen-rich
AGB stars, it has to be considered that the sample is biased. However,
recent CN Zeeman splitting observations \citep{Herpin09} seem to
indicate that similar strength fields are found around carbon-rich
stars.

Beyond determining the magnetic field strength, the large scale
structure of the magnetic field is more difficult to determine,
predominantly because the maser observations often probe only limited
line-of-sights. Even though specifically OH observations seem to
indicate a systematic field structure, it has often been suggested
that there might not be a large scale component to the field that
would be necessary to shape the outflow \citep{Soker02}. So far the
only shape constraints throughout the envelope have been determined
for the field around the supergiant star VX~Sgr (Fig.\ref{vxsgr}),
where maser observations spanning 3 orders of magnitude in distance
are all consistent with a large scale, possibly dipole shaped,
magnetic field.

\subsection{Post-AGB Stars and P-PNe}

\articlefigure[width=12cm]{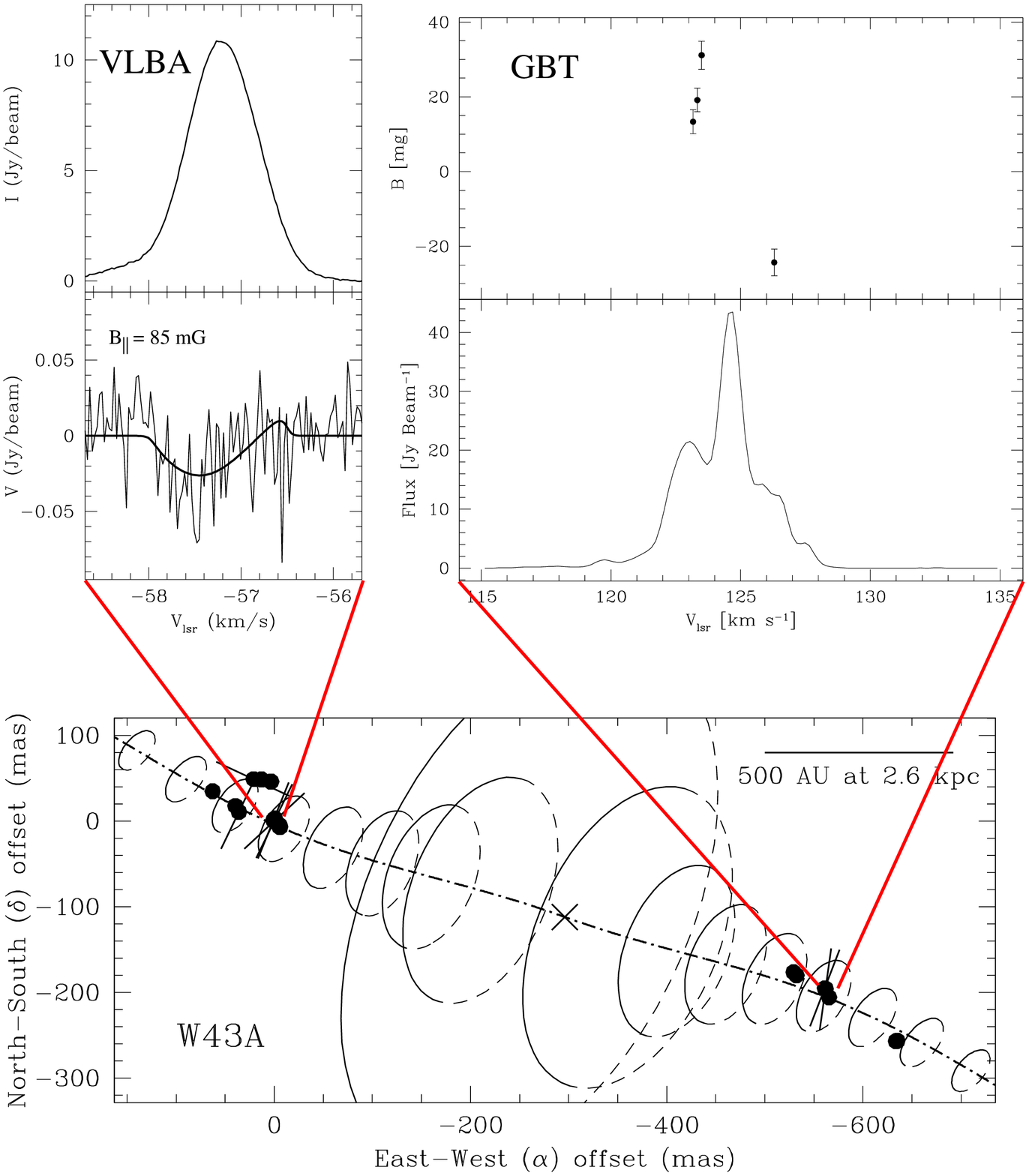}{w43a}{(top left) Total power
  (I) and V-spectrum for one of the H$_2$O maser features in
  red-shifted lobe of the collimated jet of W43A including the best
  model fit of the V-spectrum corresponding to a magnetic field
  \citep{Vlemmings06}. (top right) Confirmation of the magnetic
  field from single-dish GBT observations in the blue-shifted side of
  the lobe. As expected for being toroidal, the magnetic field
  reverses sign across the blue-shifted masers \citep{Amiri10}. (bottom) The H$_2$O masers in the precessing jet
  (dashed-dotted line) of W43A (indicated by the cross) and the
  toroidal magnetic field of W43A. The vectors indicate the determined
  magnetic field direction, perpendicular to the polarization vectors,
  at the location of the H$_2$O masers. The ellipses indicate the
  toroidal field along the jet, scaled with magnetic field strength
  $\propto r^{-1}$.}

Similar to the AGB stars, masers are the major source of magnetic
field information of post-AGB and P-PNe, with the majority of
observations focused on OH masers. These have revealed magnetic field
strengths similar to those of AGB stars (few mG) and a clear large
scale magnetic field structure \citep[e.g.][]{Bains03}.

The most promising results have come after the detection of the
so-called 'water-fountain' sources. These sources exhibit fast and
highly collimated H$_2$O maser jets that often extend beyond even the
regular OH maser shell. With the dynamical age of the jet of order 100
years, they potentially are the progenitors of the bipolar
(P-)PNe. Although the masers are often too weak for a detection of the
magnetic field, observations of the arch-type of the water-fountains,
W43A, have revealed a strong toroidal magnetic field that is
collimating the jet \citep[Fig.~\ref{w43a} and][]{Vlemmings06}.

\subsection{Planetary Nebulae}

During the PN phase, masers are rare and weak and until now only the
PN K3-35 has had a few mG magnetic field measured in its OH masers
\citep{Miranda01}. Fortunately, there are a few other methods of
measuring PN magnetic fields. The field orientation in the dust of the
nebula can be determined using dust continuum polarization
observations and current observations seem to indicate toroidal
fields, with the dust alignment likely occurring close to the dust
formation zone \citep{Sabin07}. Faraday rotation studies are
potentially also able to study the magnetic field in the interaction
region between the interstellar medium and the stellar outflow.

In contrast to AGB stars, the central stars of PNe also show atomic
lines that can be used to directly probe the magnetic fields on the
surface of these stars. While measurements are still rare,
observations of for example the central star of NGC~1360, indicate a
field of order several kG \citep{Jordan05}.

\section{Origin of the Magnetic Field}

Despite the strong observational evidence for evolved star magnetic
fields, the origin of these fields is still unclear. In single stars,
differential rotation between the AGB star core and the envelope could
potentially result in sufficiently strong magnetic field
\citep{Blackman01}. However, as the energy loss due to a rotating
magnetic field drag drains the rotation needed to maintain the field
within several tens of years, an additional source of energy is needed
\citep[e.g.][]{Nordhaus06}. If AGB stars would be able to have a
sun-like convective dynamo, magnetically dominated explosions could
indeed result from single stars. Alternatively, the energy could be
provided by the interaction with a circumstellar disk, although the
origin of the disk is then another puzzle.

Another explanation for maintaining a magnetic field is the
interaction between a binary companion or potentially a heavy planet,
with common-envelope evolution providing paths to both magnetically as
well as thermally driven outflows \citep{Nordhaus06}. A companion
could be the cause of the precession seen in a number of water-fountain
and (P-)PNe jets. However, to date, the majority of the stars with
measured magnetic fields do not show any other indication of binarity.

\begin{center}
\begin{table}
\caption{Energy densities in AGB envelopes}
\begin{tabular}{ l l || c | c | c | c }
\hline
 & & Photosphere & SiO & H$_2$O & OH \\
\hline
 & & & & & \\
 $B$ & [G] & $\sim50$? & $\sim3.5$ & $\sim0.3$ & $\sim0.003$ \\
 $R$ & [AU] & - & $\sim3$ & $\sim25$ & $\sim500$ \\
  & & - & $[2-4]$ & $[5-50]$ & $[100-10.000]$ \\
 $V_{\rm exp}$ & [km~s$^{-1}$] & $\sim5$ & $~\sim5$ & $\sim8$ & $\sim10$ \\
 $n_{\rm H_2}$ & [cm$^{-3}$] & $\sim10^{14}$ & $\sim10^{10}$& $\sim10^{8}$ & $\sim10^{6}$ \\
 $T$ & [K] & $\sim2500$ & $\sim1300$ & $\sim500$ & $\sim300$ \\
 & & & & & \\
\hline 
 & & & & & \\
 $B^2/8\pi$ & [dyne~cm$^{-2}$] & $\mathbf{10^{+2.0}}$? & $\mathbf{10^{+0.1}}$ & $\mathbf{10^{-2.4}}$ & $10^{-6.4}$ \\
 $nKT$ & [dyne~cm$^{-2}$] & $10^{+1.5}$ & $10^{-2.8}$ & $10^{-5.2}$ & $10^{-7.4}$ \\
 $\rho V_{\rm exp}^2$ & [dyne~cm$^{-2}$] & $10^{+1.5}$ & $10^{-2.5}$ & $10^{-4.1}$ & $\mathbf{10^{-5.9}}$ \\
 $V_A$ & [km~s$^{-1}$] & $\sim15$ & $\sim100$ & $\sim300$ & $\sim8$ \\
 & & & & & \\
\hline
\end{tabular}
\label{energy}
\end{table}
\end{center}

\section{Effect of the Magnetic Field}

Until a more complete sample of magnetically active AGB stars, post-AGB stars
and (P-)PNe is known, it is hard to observationally determine
the effect of the magnetic field on these late stages of
evolution. Starting with the AGB phase, a number of theoretical works
have described the potential of magnetic fields in (at least partly)
driving the stellar mass-loss through Alfv{\'e}n
waves\citep[e.g.][]{Falceta02}, or through the creation of cool spots
on the surface above with dust can form easier \citep{Soker98}. As
current models of dust and radiation driven winds are still unable to
explain especially the mass-loss of oxygen-rich stars, magnetic fields
might provide the missing component of this problem, with tentative
evidence already pointing to a relation between the magnetic field
strength and mass-loss rate.

Other theoretical works have focused on the magnetic shaping of the
stellar winds \citep[e.g.][]{Chevalier94, Garcia05, Frank04}. But to
properly determine the possible effect of the magnetic fields, it is
illustrative to study the approximate ratios of the magnetic, thermal
and kinematic energies contained in the stellar wind. In
Table.\ref{energy} I list these energies along with the Alfv{\'e}n
velocities and typical temperature, velocity and temperature parameters
in the envelope of AGB stars. While many values are quite uncertain,
as the masers that are used to probe them can exist in a fairly large
range of conditions, it seems that the magnetic energy dominates out to
$\sim50-100$~AU in the circumstellar envelope. This would correspond
to the so-called 'launch' region of magneto-hydrodynamic (MHD)
outflows, which typically extend to no more than $\sim50R_i$, with
$R_i$ the inner-most radial scale of launch engine
\citep[e.g.][]{Blackman09}. A rough constraint on $R_i$ thus seems to
be $\sim1-2$~AU, close to the surface of the star.

\section{Outlook}

While progress in studying the magnetic fields of evolved stars has
been significant, a number of crucial questions remain to be answered.
Several of these can be addressed with the new and upgraded telescopes
in the near future. For example, the upgraded EVLA and eMERLIN will
uniquely be able to determine the location of the masers in the
envelope with respect to the central star, giving us, together with
polarization observations, crucial information on the shape and
structure of the magnetic field throughout the envelopes. ALMA will be
able to add further probes of magnetic fields with for example high
frequency masers and CO polarization observations, significantly
expanding our sample of stars with magnetic field measurements. With
the ALMA sensitivity, polarization will be easily detectable even in
short observations and thus, even if not the primary goal,
polarization calibration should be done. The new low-frequency arrays
can potentially be used to determine magnetic fields in the interface
between the ISM and PNe envelopes through Faraday rotation
observations.

With the advances in the search for binaries and the theories of common-envelope evolution and MHD outflow launching, the new observations will address for example:
\begin{itemize}
\item{Under what conditions does the magnetic field dominate over e.g. binary interaction when shaping outflows?}
\item{Are magnetic fields as widespread in evolved stars as they seem?}
\item{What is the origin of the AGB magnetic field - can we find the binaries/heavy planets that might be needed?}
\item{Is there a relation between AGB mass-loss and magnetic field strength?}
\end{itemize}

\acknowledgements WV acknowledges the support by the Deutsche
Forschungsgemeinschaft (DFG) through the Emmy Noether Research grant
VL 61/3-1, and the work by the various researchers that have been
crucial in the development of the area of evolved star magnetic field
research (including those that I neglected to reference in this
review).

\bibliographystyle{asp2010}


\begin{thebibliography}{}
\expandafter\ifx\csname natexlab\endcsname\relax\def\natexlab#1{#1}\fi
\expandafter\ifx\csname url\endcsname\relax
  \def\url#1{\texttt{#1}}\fi
\expandafter\ifx\csname urlprefix\endcsname\relax\def\urlprefix{URL }\fi
\providecommand{\eprint}[2][]{\url{#2}}

\bibitem[{{Amiri} et~al.(2010){Amiri}, {Vlemmings}, \& {van
  Langevelde}}]{Amiri10}
{Amiri}, N., {Vlemmings}, W., \& {van Langevelde}, H.~J. 2010, \aap, 509, A26+

\bibitem[{{Bains} et~al.(2003){Bains}, {Gledhill}, {Yates}, \&
  {Richards}}]{Bains03}
{Bains}, I., {Gledhill}, T.~M., {Yates}, J.~A., \& {Richards}, A.~M.~S. 2003,
  \mnras, 338, 287

\bibitem[{{Balick} \& {Frank}(2002)}]{Balick02}
{Balick}, B., \& {Frank}, A. 2002, \araa, 40, 439

\bibitem[{{Blackman}(2009)}]{Blackman09}
{Blackman}, E.~G. 2009, in IAU Symposium, vol. 259 of IAU Symposium, 35

\bibitem[{{Blackman} et~al.(2001){Blackman}, {Frank}, {Markiel}, {Thomas}, \&
  {Van Horn}}]{Blackman01}
{Blackman}, E.~G., {Frank}, A., {Markiel}, J.~A., {Thomas}, J.~H., \& {Van
  Horn}, H.~M. 2001, \nat, 409, 485

\bibitem[{{Bujarrabal} et~al.(2001){Bujarrabal}, {Castro-Carrizo}, {Alcolea},
  \& {S{\'a}nchez Contreras}}]{Bujarrabal01}
{Bujarrabal}, V., {Castro-Carrizo}, A., {Alcolea}, J., \& {S{\'a}nchez
  Contreras}, C. 2001, \aap, 377, 868

\bibitem[{{Chevalier} \& {Luo}(1994)}]{Chevalier94}
{Chevalier}, R.~A., \& {Luo}, D. 1994, \apj, 421, 225

\bibitem[{{Falceta-Gon{\c c}alves} \& {Jatenco-Pereira}(2002)}]{Falceta02}
{Falceta-Gon{\c c}alves}, D., \& {Jatenco-Pereira}, V. 2002, \apj, 576, 976

\bibitem[{{Frank} \& {Blackman}(2004)}]{Frank04}
{Frank}, A., \& {Blackman}, E.~G. 2004, \apj, 614, 737

\bibitem[{{Garc{\'{\i}}a-Segura} et~al.(2005){Garc{\'{\i}}a-Segura},
  {L{\'o}pez}, \& {Franco}}]{Garcia05}
{Garc{\'{\i}}a-Segura}, G., {L{\'o}pez}, J.~A., \& {Franco}, J. 2005, \apj,
  618, 919

\bibitem[{{Herpin} et~al.(2006){Herpin}, {Baudry}, {Thum}, {Morris}, \&
  {Wiesemeyer}}]{Herpin06}
{Herpin}, F., {Baudry}, A., {Thum}, C., {Morris}, D., \& {Wiesemeyer}, H. 2006,
  \aap, 450, 667

\bibitem[{{Herpin} et~al.(2009){Herpin}, {Baudy}, {Josselin}, {Thum}, \&
  {Wiesemeyer}}]{Herpin09}
{Herpin}, F., {Baudy}, A., {Josselin}, E., {Thum}, C., \& {Wiesemeyer}, H.
  2009, in IAU Symposium, vol. 259 of IAU Symposium, 47

\bibitem[{{Jordan} et~al.(2005){Jordan}, {Werner}, \& {O'Toole}}]{Jordan05}
{Jordan}, S., {Werner}, K., \& {O'Toole}, S.~J. 2005, \aap, 432, 273

\bibitem[{{Kemball} et~al.(2009){Kemball}, {Diamond}, {Gonidakis}, {Mitra},
  {Yim}, {Pan}, \& {Chiang}}]{Kemball09}
{Kemball}, A.~J., {Diamond}, P.~J., {Gonidakis}, I., {Mitra}, M., {Yim}, K.,
  {Pan}, K., \& {Chiang}, H. 2009, \apj, 698, 1721

\bibitem[{{Kwok} et~al.(1978){Kwok}, {Purton}, \& {Fitzgerald}}]{Kwok78}
{Kwok}, S., {Purton}, C.~R., \& {Fitzgerald}, P.~M. 1978, \apjl, 219, L125

\bibitem[{{Miranda} et~al.(2001){Miranda}, {G{\'o}mez}, {Anglada}, \&
  {Torrelles}}]{Miranda01}
{Miranda}, L.~F., {G{\'o}mez}, Y., {Anglada}, G., \& {Torrelles}, J.~M. 2001,
  \nat, 414, 284

\bibitem[{{Nordhaus} \& {Blackman}(2006)}]{Nordhaus06}
{Nordhaus}, J., \& {Blackman}, E.~G. 2006, \mnras, 370, 2004

\bibitem[{{Rudnitski} et~al.(2010){Rudnitski}, {Pashchenko}, \&
  {Colom}}]{Rudnitski10}
{Rudnitski}, G.~M., {Pashchenko}, M.~I., \& {Colom}, P. 2010, Astronomy
  Reports, 54, 400

\bibitem[{{Sabin} et~al.(2007){Sabin}, {Zijlstra}, \& {Greaves}}]{Sabin07}
{Sabin}, L., {Zijlstra}, A.~A., \& {Greaves}, J.~S. 2007, \mnras, 376, 378

\bibitem[{{Soker}(1998)}]{Soker98}
{Soker}, N. 1998, \mnras, 299, 1242

\bibitem[{{Soker}(2002)}]{Soker02}
--- 2002, \mnras, 336, 826

\bibitem[{{Szymczak} et~al.(2001){Szymczak}, {Cohen}, \&
  {Richards}}]{Szymczak01}
{Szymczak}, M., {Cohen}, R.~J., \& {Richards}, A.~M.~S. 2001, \aap, 371, 1012

\bibitem[{{Vlemmings}(2007)}]{Vlemmings07}
{Vlemmings}, W.~H.~T. 2007, in IAU Symposium, edited by {J.~M.~Chapman \&
  W.~A.~Baan}, vol. 242 of IAU Symposium, 37

\bibitem[{{Vlemmings} et~al.(2006){Vlemmings}, {Diamond}, \&
  {Imai}}]{Vlemmings06}
{Vlemmings}, W.~H.~T., {Diamond}, P.~J., \& {Imai}, H. 2006, \nat, 440, 58

\bibitem[{{Vlemmings} et~al.(2002){Vlemmings}, {Diamond}, \& {van
  Langevelde}}]{Vlemmings02}
{Vlemmings}, W.~H.~T., {Diamond}, P.~J., \& {van Langevelde}, H.~J. 2002, \aap,
  394, 589

\bibitem[{{Vlemmings} et~al.(2010){Vlemmings}, {Humphreys}, \& {Franco-Hern{\'a}ndez}}]{Vlemmings10}
{Vlemmings}, W.~H.~T., {Humphreys}. E.~M.~L., \& {Franco-Hern{\'a}ndez}, R. 2010, \apj, submitted

\bibitem[{{Vlemmings} et~al.(2005){Vlemmings}, {van Langevelde}, \&
  {Diamond}}]{Vlemmings05}
{Vlemmings}, W.~H.~T., {van Langevelde}, H.~J., \& {Diamond}, P.~J. 2005, \aap,
  434, 1029

\end{thebibliography}

\end{document}